\documentclass[adp,fleqn]{w-art}
\usepackage{times}
\usepackage{w-thm}
\usepackage[]{graphicx}
\usepackage[breaklinks=true,colorlinks=true,linkcolor=blue,citecolor=red,filecolor=blue,urlcolor=blue]{hyperref}
\usepackage{psfrag}
\begin{document}
\DOIsuffix{theDOIsuffix}
\Volume{16}
\Issue{1}
\Copyrightissue{01}
\Month{01}
\Year{2007}
\pagespan{3}{}
\keywords{Cosmological phase transitions, inflation, initial conditions and eternal universe, cosmology.}
\subjclass[pacs]{11.27.+d, 95.30.Sf, 98.80.-k}
\title[Vacuum bubbles on matter backgrounds]{On the fate of vacuum bubbles on matter backgrounds}
\author[A. Raki\'c]{Aleksandar Raki\'c\inst{1,}%
\footnote{Corresponding author\quad E-mail:~\textsf{rakic@astro.uni-wuerzburg.de}, Phone: +49\,931\,32\,83676, Fax: +49\,931\,888\,4603}}
\address[\inst{1}]{ITPA Universit\"at W\"urzburg, Am Hubland, 97074 W\"urzburg, Germany}
\author[D. Simon]{Dennis Simon\inst{1,}
\footnote{E-mail:~\textsf{dsimon@astro.uni-wuerzburg.de} }}
\author[J. Adamek]{Julian Adamek\inst{1,}
\footnote{E-mail:~\textsf{jadamek@astro.uni-wuerzburg.de} }}
\author[J. Niemeyer]{Jens C. Niemeyer\inst{1,2,}%
\footnote{E-mail:~\textsf{niemeyer@astro.physik.uni-goettingen.de} }}
\address[\inst{2}]{Institut f\"ur Astrophysik, Universit\"at G\"ottingen, Friedrich-Hund-Platz 1, 37077 G\"ottingen, Germany}
\begin{abstract}
In this letter we discuss cosmological first order phase transitions with de Sitter bubbles nucleating on (inhomogeneous) matter backgrounds. The de Sitter bubble can be a toy model for an inflationary phase of universes like our own. Using the thin wall approximation and the Israel junction method we trace the classical evolution of the formed bubbles within a compound model. We first address homogeneous ambient space (FRW model) and already find that bubbles nucleated in a dust dominated background cannot expand. For an inhomogeneous dust background (LTB model) we describe cases with at least initially expanding bubbles. Yet, an ensuing passage of the bubble wall through ambient curvature inhomogeneities remains unnoticed for observers inside the bubble. Notable effects also for interior observers are found in the case of a rapid background phase transition in a FRW model.
\end{abstract}

\maketitle

\newcommand{\rmd}{\rm d}

\section{Introduction}
The influence of background inhomogeneity can for instance be important in the context of string landscape sampling via tunneling processes. Standardly, bubble evolution has been studied almost exclusively in de Sitter spacetime, see e.g.~\cite{Aguirre05}, given the long lifetimes of the metastable vacua in the landscape and the cosmic no hair conjecture. However, there are recent scenarios in which tunneling is catalysed and can be made very \emph{rapid}, such as chain inflation \cite{Freese05} or DBI and resonance tunneling \cite{Sarangi07,Tye06}. Therefore we consider bubble propagation on backgrounds that contain homogeneously and inhomogeneously distributed dust. Inhomogeneous initial states could be relevant especially for processes of resonant tunneling, see \cite{Saffin08,Copeland07} and \cite{Tye09}. Another context is the question of whether and how non-standard backgrounds influence the onset of inflation, see e.g.~\cite{Goldwirth91}. More details on the methodology, more numerical examples and a perspective on the tunneling process on time dependent backgrounds can be found in \cite{Simon09}.

\section{Junction Method}
The bubble wall, which we assume to be thin w.r.t.~the bubble size, separates interior and exterior spacetimes. In order to analyse the effects of a matter background on the bubble evolution we make use of the Israel junction conditions \cite{Israel}, similar to \cite{Fischler07}. The interior spacetime is fixed to be de Sitter and can be a toy model for the inflation of our own universe. We describe the bubble wall by the induced metric $ h_{ij} {\rmd}y^i {\rmd}y^j \equiv -{\rmd\tau}^2 + R^2{\rmd\Omega}^2$ and stress-energy $S_{ij} = -\sigma h_{ij}$, with bubble surface tension $\sigma$. The non-standard part is the exterior which is modelled by a Lema\^ itre-Tolman-Bondi (LTB) model \cite{Lemaitre}:
\begin{equation} 
  {\rmd}s^2 = -{\rmd}t^2 + \frac{\left(r\partial_r a(t,r) + a(t,r)\right)^2}{1+2E(r)}{\rmd}r^2 + a^2(t,r)r^2{\rmd}\Omega^2 \;.
\end{equation}
The LTB spacetime is spherically symmetric but radially inhomogeneous. The function $E(r)$ is the local curvature of constant time slices. With dust plus vacuum source, two non-trivial field equations remain
\begin{equation}
  \left(\frac{\partial_t a}{a}\right)^2 - \frac{2E}{a^2r^2} = \frac{2M}{a^3r^3} +\frac{\Lambda}{3} \quad \text{and} \quad 8\pi\rho = \frac{2\partial_r M}{a^2r^2\left(r\partial_ra+a\right)} \;,
\end{equation}
There are three free functions in the model: $M(r)$, $E(r)$ and the bang time. At the same time there is a remaining gauge freedom in $r$. For $\partial_r M >0$ one can rescale $r$ such that $M(r) = \frac{4\pi}{3}A r^3$, with $A$ a constant. So, the LTB model leaves us with two free functions: the local curvature and the bang time.

In order to glue the spacetimes together we use the method of Israel \cite{Israel}. This implies two junction conditions for the discontinuity of the induced metric and extrinsic curvature on the bubble wall $\Sigma$:
\begin{equation} 
 \left[h_{ij}\right] \equiv h_{ij}^+\vert_\Sigma -h_{ij}^-\vert_\Sigma = 0  \quad \text{and} \quad  \left[K_{ij}\right] - h_{ij}\left[K\right] = 8\pi S_{ij} \;,
\end{equation}
with $\pm$ labels denoting exterior/interior spacetime. Note that the second equation is closely related to a continuity equation of the form $\nabla_i S_j^i + [T^\alpha_\beta n_\alpha e^\beta_j] = 0$, where $n^\mu$ is a normal vector on $\Sigma$. We consider the exterior stress-energy as given by a perfect fluid and the interior one by a cosmological constant.

From the junction conditions we get the final equations of motion in the LTB frame, they read
\begin{equation}
 \partial_t \bar r = \frac{-(1+2E)\bar r\partial_t a +\sqrt{(1+2E)\left(1+2V\right)Y}}{\left(\bar r\partial_{\bar r} a+a\right)\left(2E-2V\right)} \;, \;
 \partial_t \sigma = \rho\frac{\left(\bar r\partial_{\bar r} a+a\right)\partial_t \bar r} {\sqrt{1+2E-\left(\bar r\partial_{\bar r} a+a\right)^2\left(\partial_t \bar r\right)^2}}
\label{ijs}
\end{equation}
with $\bar r(t)$ the bubble trajectory in LTB coordinates and $Y=(\bar r\partial_t a)^2-2E+2V$. The potential $V$ reads
\begin{equation}
 2V =  -\left[\frac{\Lambda_-}{3} + \left(\frac{A}{3a^3\sigma} +\frac{\Lambda_+ -\Lambda_-}{24\pi\sigma} +2\pi\sigma \right)^2\right]R^2 \quad \text{with} \quad
4\pi\sigma < \sqrt{\frac{8\pi A}{3a^3} + \frac{\Lambda_+ -\Lambda_-}{3}}
\label{constraint}
\end{equation}
being an additional and purely geometrical constraint for the surface tension that stems from our choice of fixing the sign of the normal vector on $\Sigma$.

\begin{figure}[ht]
\centering
\psfrag{r/e}[b][b]{\small $\rho / \epsilon_{\mathrm{vac}}$}
\psfrag{6pGs2/e}[t][t]{\small $6 \pi \sigma^2 / \epsilon_{\mathrm{vac}}$}
\psfrag{collapsing}[b][b]{\small \textbf{contracting}}
\psfrag{expanding}[t][t]{\small \textbf{expanding}}
\psfrag{forbidden}[b][b]{\small \textbf{forbidden}}
\psfrag{smallest possible rL}[b][b]{\footnotesize matter dominated universes possible}
\psfrag{vacuum dominated universes}[t][t]{\footnotesize all universes vacuum dominated}
\includegraphics[width=0.44\textwidth]{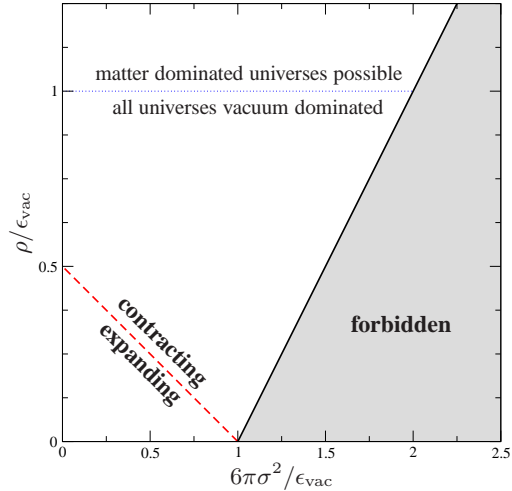}
\caption{\label{RhoVsSigm}
The early-time behaviour of comovingly nucleated de Sitter bubbles on FRW background
with vacuum as well as dust source. The fate of the bubble as seen by an exterior 
comoving observer crucially depends on how the dust density compares to the surface tension of the bubble $\sigma$ and the latent heat of the vacuum $\epsilon_{\mathrm{vac}} \equiv \left(\Lambda_+ - \Lambda_-\right) / 8 \pi$, see also eq.~(\ref{budget}). In order to to produce an expanding bubble only little dust density in the background is allowed -- on matter dominated backgrounds ($\rho > \rho_{\mathrm{vac}} = \Lambda_+ / 8 \pi$) the nucleated bubbles must collapse. The shaded region in the plot is excluded by the junction constraint (\ref{constraint}).}
\end{figure}

First, let us consider the \emph{homogeneous case}. Already the study of this case can put restrictions on the applicability of the described junction method and will throw off some interesting physics. In addition to $\Lambda_\pm$ we put an amount of dust density in the outside space. Its initial distribution is assumed to be homogeneous and local curvature is assumed to vanish: this is the \emph{homogeneous limit} of the LTB model which reduces to a FRW model. The fate of a bubble of new vacuum that nucleates comovingly ($\partial_t \bar{r} = 0$) on such a background will depend crucially on the force balance and thus on the relation of the latent heat to surface tension and dust density. After differentiating (\ref{ijs}) we have
\begin{equation}
\left. \partial_t^2 \bar{r}\right|_{\partial_t \bar{r} = 0}~=~\frac{1}{a} \left(\frac{\Lambda_+ - \Lambda_-}{24 \pi \sigma} - 2 \pi \sigma - \frac{2 \rho}{3 \sigma}\right) \; .
\label{budget}
\end{equation}
In contrast to nucleation on a pure vacuum background the presence of matter can now impede the expansion of a vacuum bubble. There is a competition between the surface tension that tries to collapse the bubble and the pressure difference from the latent heat of the vacuum. In the pure vacuum case the bubble wall will move in the direction of smaller pressure, i.e.~into the space of higher vacuum energy. As soon as a non-negligible contribution of pressureless dust is added to the budget the pressure support cannot prevail anymore and bubbles would allways collapse. Only for vacuum dominated setups ($\rho < \rho_{\mathrm{vac}} \equiv \Lambda_+ / 8 \pi$, and $\Lambda_+ > \Lambda_- \geq 0$) expanding bubbles can exist. In a dust dominated background the initial acceleration of the bubble becomes negative leading to bubble contraction. The situation is illustrated in Fig.~\ref{RhoVsSigm}.

\begin{figure}
\begin{center}
\begin{tabular}{lr}
  \psfrag{Hr}[][]{\scriptsize{$\sqrt{\Lambda_+/3} r$}}
  \psfrag{y}[][]{\scriptsize{$R_\mathrm{cr}\sqrt{k}$}}
  \includegraphics[width=0.42\textwidth]{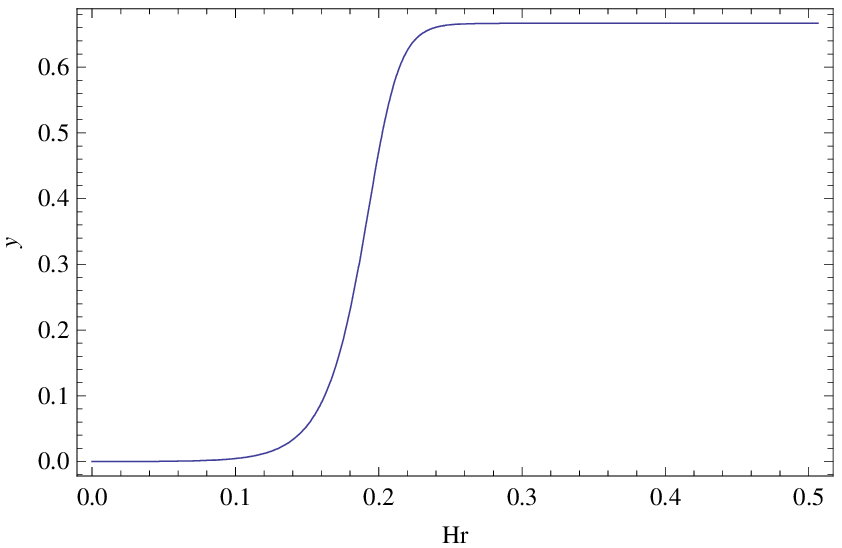}
&
  \psfrag{y}[][]{\scriptsize{$\sqrt{\Lambda_-/3}~\bar r$}}
  \psfrag{x}[][]{\scriptsize{$\sqrt{\Lambda_-/3}~t$}}
  \includegraphics[width=0.42\textwidth]{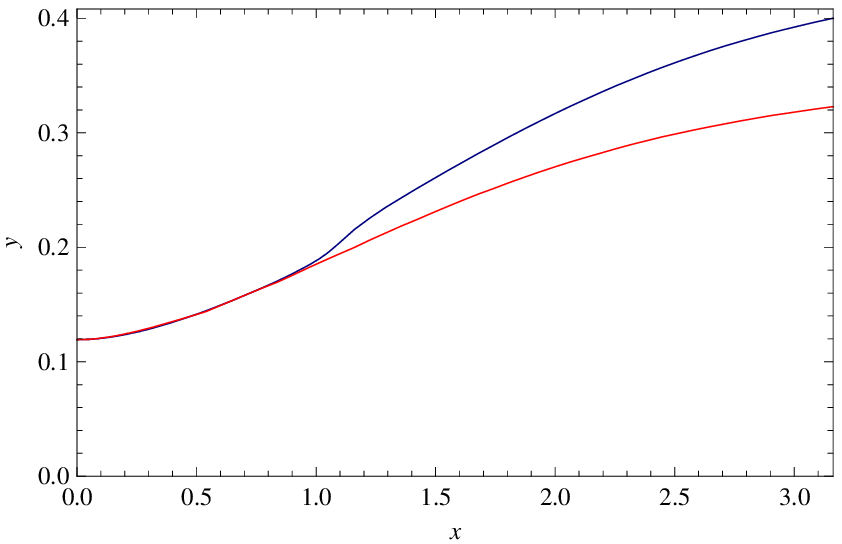}\\

  \psfrag{y}[][]{\scriptsize{$\sqrt{\Lambda_+/3}~\bar r$}}
  \psfrag{x}[][]{\scriptsize{$\sqrt{\Lambda_+/3}~t$}}
  \includegraphics[width=0.42\textwidth]{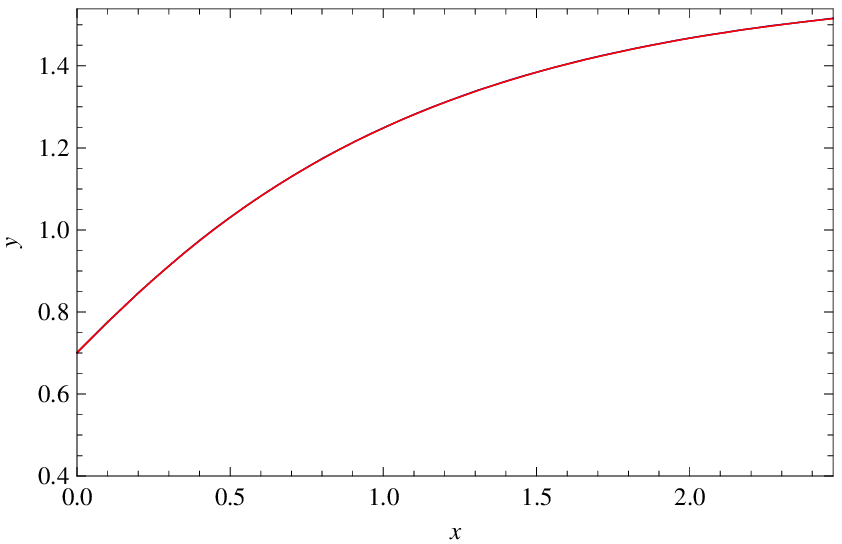}
&
  \psfrag{y}[][]{\scriptsize{$\sigma/\sigma_0$}}
  \psfrag{x}[][]{\scriptsize{$\sqrt{\Lambda_-/3}~t$}}
  \includegraphics[width=0.42\textwidth]{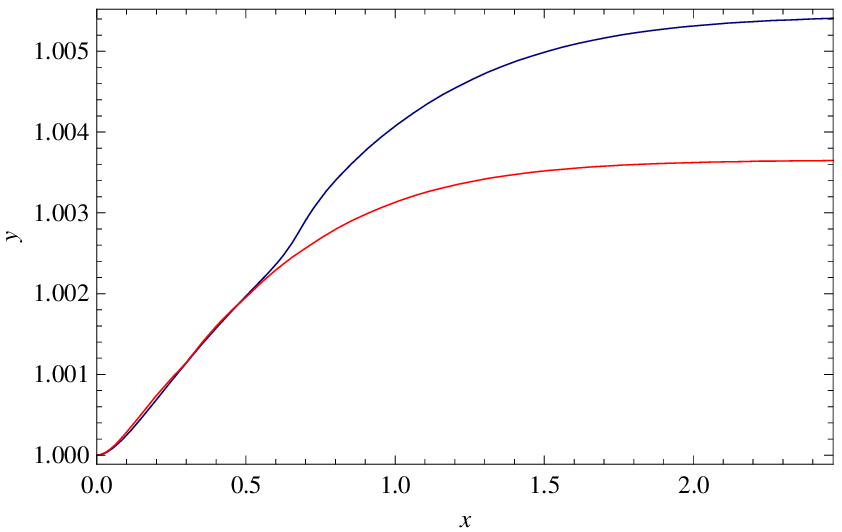}
\end{tabular}
  \caption{\label{plot2}
The numerical evolution of a growing de Sitter bubble on an LTB background which is inhomogeneous in its initial dust profile \emph{and} has a curvature inhomogeneity. The profile of the latter is shown in the upper left figure: its form is chosen as a tanh, its amplitude is bounded by shell crossing. The upper right figure compares the trajectories of the vacuum bubble on an undisturbed flat background (red) with such on the LTB background (blue) as seen from the (exterior) LTB frame. The distinct perturbation from the curvature disturbance seen in this frame vanishes when the trajectory is computed in de Sitter coordinates (lower left figure). The effect is a coordinate artefact -- nevertheless, considerable fine-tuning is needed in order to be even able to study such transitions within the junction method. As a consequence of the junction approach there is a notable effect on the surface tension of the bubble (lower right figure).}
\end{center}
\end{figure}

Now let us consider the \emph{inhomogeneous LTB case}. There are two functions with which we can introduce inhomogeneity: the local curvature and the bang time. A choice of the latter is equivalent to a choice of the initial dust density profile $\rho_0(r)\equiv \rho(t_0,r)$. Thus we have $E(r)$ and $\rho_0(r)$ at our disposal. What we are looking for in the end is whether the motion of a bubble through an inhomogeneity leads to potentially observable disturbances in the trajectory of the wall. The result of the last section shows that the presence of dust has significant influence on the trajectory and makes it difficult for the bubble wall to propagate to an exterior inhomogeneity at all. Note that also relaxing the condition of a comoving nucleation and considering bubbles with  $\partial_t \bar r(t_0)>0$ does not save the bubbles from collapsing whenever there is enough dust in the background. However, we can enforce a transition by combining both free LTB functions. First we take an inhomogeneous initial dust distribution with a radially increasing profile -- then small bubbles nucleate in a still vacuum dominated region and can initially expand. Second we take a curvature inhomogeneity and place it right in the way of the expanding bubble wall -- note that the curvature profile is constrained due to shell crossing. For this twofold inhomogeneous setup we have compared the evolution of the bubble wall trajectory with its evolution for the homogeneous case and found a difference only in outside coordinates. In the (more relevant) inside frame the transition remains unnoticed, cf. Fig.~\ref{plot2}.

We also considered an \emph{FRW background} with a perfect fluid that is just undergoing a \emph{phase transition}, say a reheating or vice versa phase transition, while de Sitter bubbles evolve on it \cite{Simon09}. In contrast to the dust case the phase transition leaves a distinct perturbation of the wall trajectory, also seen in the inside.

\section{Conclusions}
On a de Sitter background vacuum bubbles generically expand into the space of higher vacuum energy. Here we have seen that a dust dominated homogeneous background hinders the expansion of de Sitter bubbles. Consequently, vacuum bubbles can hardly propagate towards given matter inhomogeneities in the background. In the LTB model we have used for the inhomogeneous background we have two free functions at disposal: the initial dust density profile and the spatial curvature profile. For a special choice of these functions the bubble wall reaches the inhomogeneities but these do not affect the trajectory of the wall but merely the surface tension, as seen by interior (physical) observers. Nevertheless, we found that also the interior observers can potentially see a significant perturbation of the bubble wall if the curvature inhomogeneity in the ambient LTB model is replaced by a smooth phase transition, say $w=1/3 \rightarrow w=-1$ or vice versa, taking place in a FRW background, cf. \cite{Simon09}.

If such disturbances of our bubble wall do exist it might not be impossible to observe them. In \cite{Aguirre08,Chang08} a different mechanism -- primordial bubble collisions -- has been discussed and it is argued that perturbations in the bubble trajectory can actually lead to a characteristic modification in the redshift of the reheating surface and thus could potentially be detectable in the CMB.


\begin{acknowledgement}
This work was supported by the DFG through the W\"urzburg Research Training Group 1147.
\end{acknowledgement}


\begin{thebibliography}{99}

\bibitem{Aguirre05}
  A.~Aguirre and M.~C.~Johnson, 
  Phys.\ Rev.\  D {\bf 72} (2005) 103525,
  arXiv:gr-qc/0508093.


\bibitem{Freese05}
  K.~Freese and D.~Spolyar, 
  JCAP {\bf 0507} (2005) 007,
  arXiv:hep-ph/0412145.


\bibitem{Sarangi07}
  S.~Sarangi, G.~Shiu and B.~Shlaer, 
  Int.\ J.\ Mod.\ Phys.\  A {\bf 24} (2009) 741,
  arXiv:0708.4375 [hep-th].


\bibitem{Tye06}
  S.~H.~Henry Tye, 
  arXiv:hep-th/0611148.


\bibitem{Saffin08}
  P.~M.~Saffin, A.~Padilla and E.~J.~Copeland, 
  JHEP {\bf 0809} (2008) 055,
  arXiv:0804.3801 [hep-th].


\bibitem{Copeland07}
  E.~J.~Copeland, A.~Padilla and P.~M.~Saffin, 
  JHEP {\bf 0801} (2008) 066,
  arXiv:0709.0261 [hep-th].


\bibitem{Tye09}
  S.~H.~Tye and D.~Wohns, 
  arXiv:0910.1088 [hep-th].


\bibitem{Goldwirth91}
  D.~S.~Goldwirth and T.~Piran, 
  Phys.\ Rept.\  {\bf 214} (1992) 223.


\bibitem{Simon09}
  D.~Simon, J.~Adamek, A.~Raki\'c and J.~C.~Niemeyer, 
  JCAP {\bf 0911} (2009) 008,
  arXiv:0908.2757 [gr-qc].


\bibitem{Israel}
  W.~Israel, 
  Nuovo Cim.B{\bf 44S10} (1966)1, Erratum-ibid.\  B {\bf 48} (1967\ NUCIA,B44,1.1966) 463.


\bibitem{Fischler07}
  W.~Fischler, S.~Paban, M.~\v Zani\'c and C.~Krishnan, 
  JHEP {\bf 0805} (2008) 041,
  arXiv:0711.3417 [hep-th].


\bibitem{Lemaitre}
  G.~A.~Lema\^itre and M.~A.~H.~MacCallum, 
  Gen.\ Rel.\ Grav.\  {\bf 29} (1997) 641.






\bibitem{Aguirre08}
  A.~Aguirre, M.~C.~Johnson and M.~Tysanner, 
  Phys.\ Rev.\  D {\bf 79} (2009) 123514,
  arXiv:0811.0866 [hep-th].


\bibitem{Chang08}
  S.~Chang, M.~Kleban and T.~S.~Levi, 
  JCAP {\bf 0904} (2009) 025,
  arXiv:0810.5128 [hep-th].


\end{thebibliography}
\end{document}